\documentclass[]{spie}  

\usepackage{amsmath,amsfonts,amssymb}
\usepackage{graphicx}
\newcommand{\simle}{\mbox{$\stackrel{<}{_{\sim}}$}}

\title{Architecture design study and technology roadmap for the Planet Formation Imager (PFI) \footnote{Copyright 2016 Society of Photo-Optical Instrumentation Engineers. One print or electronic copy may be made for personal use only. Systematic reproduction and distribution, duplication of any material in this paper for a fee or for commercial purposes, or modification of the content of the paper are prohibited. DOI: http://dx.doi.org/10.1117/12.2233311}}

\author[a]{John D. Monnier}
\author[b]{Michael J. Ireland}
\author[c]{Stefan Kraus}
\author[d]{Fabien Baron}
\author[e]{Michelle Creech-Eakman}
\author[f]{Ruobing Dong}
\author[g]{Andrea Isella}
\author[h]{Antoine Merand}
\author[i]{Ernest Michael}
\author[j]{Stefano Minardi}
\author[k]{David Mozurkewich}
\author[l]{Romain Petrov}
\author[m]{Stephen Rinehard}
\author[n]{Theo ten Brummelaar}
\author[o]{Gautum Vasisht}
\author[p]{Ed Wishnow}
\author[q]{John Young}
\author[r]{Zhaohuan Zhu}
\affil[a]{University of Michigan, USA}
\affil[b]{Research School of Astronomy and Astrophysics, Australian National University, Canberra, ACT 2611, Australia}
\affil[c]{University of Exeter, UK}
\affil[d]{Georgia State University, USA}
\affil[e]{New Mexico Tech, USA}
\affil[f]{University of California, Berkeley, USA}
\affil[g]{Rice University, USA}
\affil[h]{European Southern Observatory, Chile}
\affil[i]{University of Chile}
\affil[j]{University of Jena, Germany}
\affil[k]{Seabrook Engineering, USA}
\affil[l]{University of Nice, France}
\affil[m]{NASA-GSFC, USA}
\affil[n]{CHARA Array, Georgia State University, USA}
\affil[o]{Jet Propulsion Laboratory, USA}
\affil[p]{University of California at Berkeley, USA}
\affil[q]{University of Cambridge, UK}
\affil[r]{Princeton University, USA}

\authorinfo{Further author information: (Send correspondence to J.D.M.)\\J.D.M.: E-mail: monnier@umich.edu, Telephone: 1 734 763 5822}

\begin{document} 
\maketitle

\begin{abstract}
The Planet Formation Imager (PFI) Project has formed a Technical Working Group (TWG) to explore possible facility architectures to meet the primary PFI science goal of imaging planet formation {\em in situ} in nearby star-forming regions.   The goals of being sensitive to dust emission on solar system scales and resolving  the Hill-sphere around forming giant planets can best be accomplished through sub-milliarcsecond imaging in the thermal infrared.  Exploiting the 8-13 micron atmospheric window, a ground-based long-baseline interferometer with approximately 20 apertures including 10km baselines will have the necessary resolution to image structure down  0.1 milliarcseconds (0.014 AU) for T Tauri disks in Taurus.  Even with large telescopes, this array will not have the sensitivity to directly track fringes in the mid-infrared for our prime targets and a fringe tracking system will be necessary in the near-infrared.    While a heterodyne architecture using modern mid-IR laser comb technology remains a competitive option (especially for the intriguing 24 and 40$\mu$m atmospheric windows), the prioritization of 3-5$\mu$m observations of CO/H$_2$O vibrotational levels by the PFI-Science Working Group (SWG) pushes the TWG to require vacuum pipe beam transport with potentially cooled optics.  We present here a preliminary study of simulated L- and N-band PFI observations of a realistic 4-planet disk simulation, finding 21x2.5m PFI can easily detect the accreting protoplanets in both L and N-band but can see non-accreting planets only in L band. We also find that even an ambitious PFI will lack sufficient surface brightness sensitivity to image details of the fainter emission from dust structures beyond $\sim$5 AU, unless directly illuminated or heated by local energy sources.  That said, the utility of PFI at N-band is highly dependent on the stage of planet formation in the disk and we require additional systematic studies in conjunction with the PFI-SWG to better understand the science capabilities of PFI, including the potential to resolve protoplanetary disks in emission lines to measure planet masses using position-velocity diagrams. We advocate for a specific technology road map in order to reduce the current cost driver (telescopes) and to validate high accuracy fringe tracking and high dynamic range imaging at L, M band.  In conclusion, no technology show-stoppers have been identified for PFI to date, however there is high potential for breakthroughs in medium-aperture (4-m class) telescopes architecture that could  reduce the cost of PFI by a factor of 2 or more.
\end{abstract}

\keywords{interferometry, mid-infrared, exoplanets, planet formation, astronomy, facilities, imaging, infrared}

\section{Introduction}
The Planet Formation Imager Project (www.planetformationimager.org) was initiated in 2014 to study the prospects for a long-baseline infrared interferometer to achieve breakthrough insights into the complex processes that control planet formation.  
We presented the first science and technical cases at the Montreal SPIE meeting\cite{monnier2014spie,kraus2014spie,ireland2014spie}.  Here we give a progress report on the workings of the PFI Technical Working Group (TWG), building upon the early work outlined in the last SPIE.
A report from the science working group can also be found in these proceedings\cite{kra16}.

\section{Science Requirements}

\begin{figure} [ht]
   \begin{center}
   \begin{tabular}{c} 
   \includegraphics[width=3.75in]{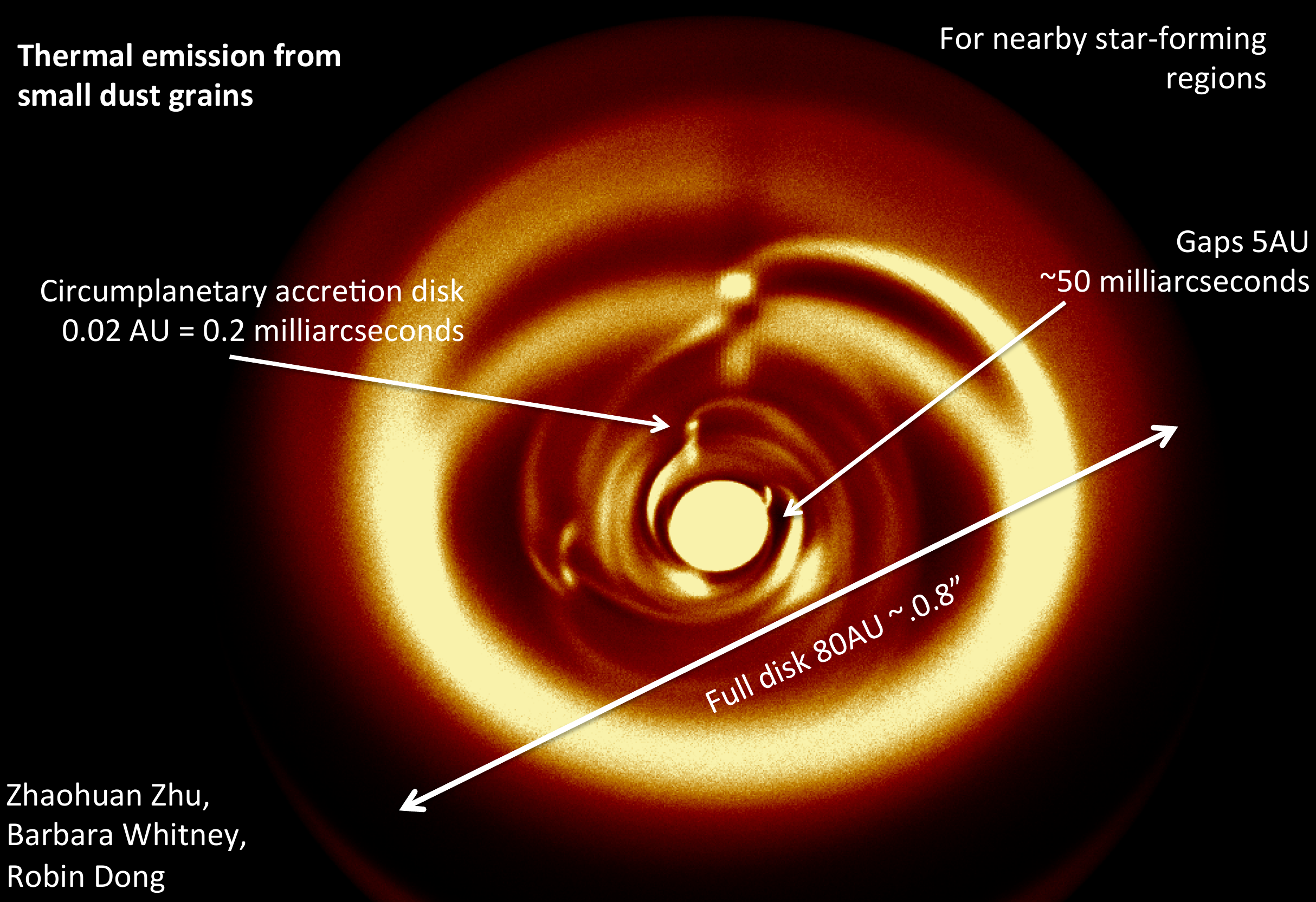}
    \includegraphics[width=2.8in]{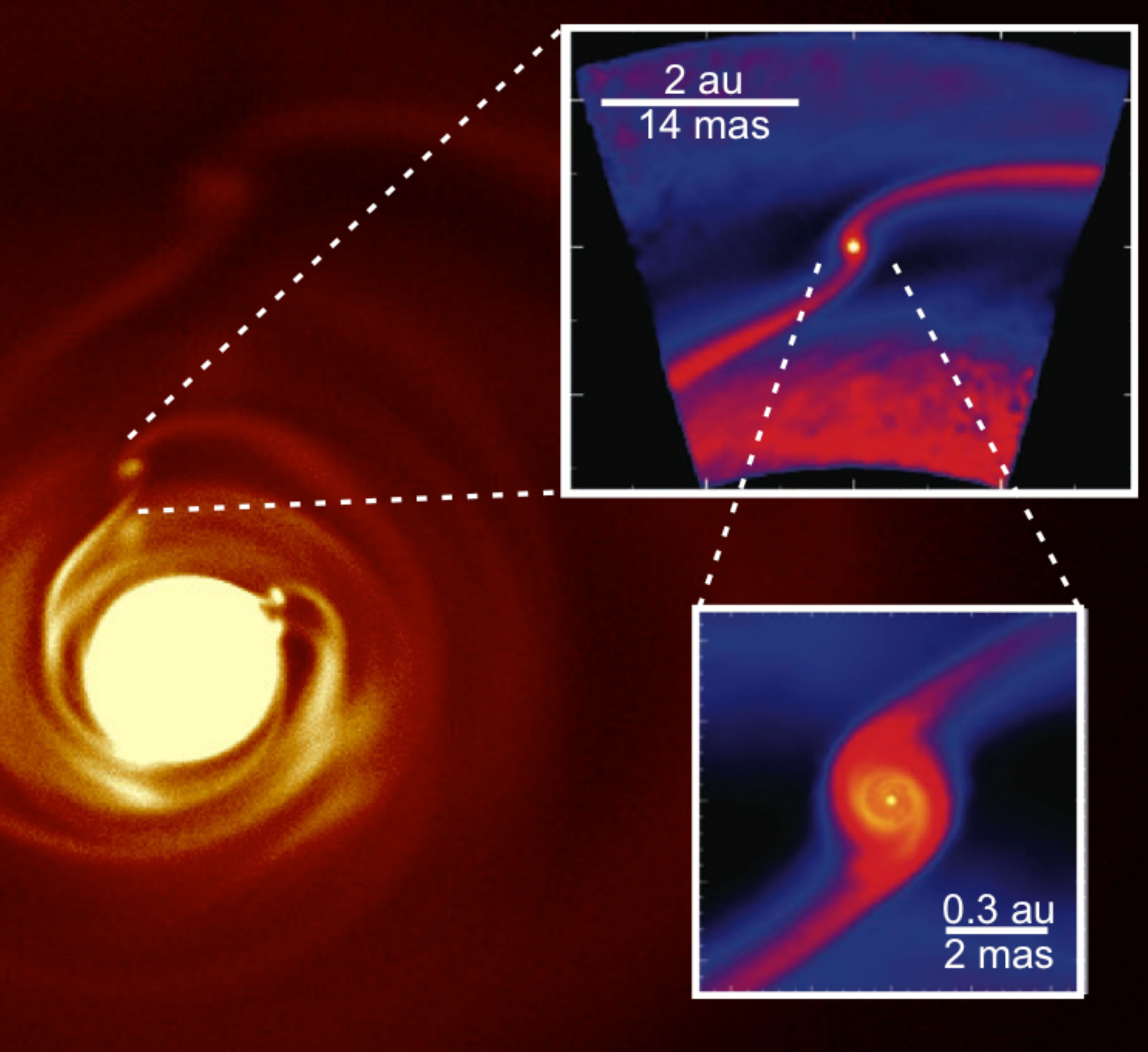}
	\end{tabular}
	\end{center}
   \caption[example] 
   { \label{fig:diskscales} 
Here we see an image of thermal emission from a planet formation simulation by Zhu et al\cite{zhu2011}.  We label the different spatial scales that are relevant for PFI, with special attention to the maximum size of protoplanetary accretion disks (Hill Sphere), which places a requirement of sub-milliarcsecond angular resolution for PFI.
On the right, we imagine zooming in one of the circumplanetary disks in order to see structures in the accretion flow (Ayliffe et al.\cite{ayliffe2009})}
   \end{figure}

The scientific goals of PFI are illustrated in Figure~\ref{fig:diskscales}, where we show a simulated image of a young star and its disk during vigorous planet formation (Zhu et al.\cite{zhu2011}). We see that mid-infrared emission from nearby young stars and their disks can subtend nearly an arcsecond, with gap structures on the scale of 0.05'' and circumplanetary accretion disks with size $\simle$0.5 milliarcseconds.  This latter scale derives from the radius of the Hill Sphere $R_H$ which depends on the mass of the planet $M_p$ and the distance $a$ from the central star\cite{Hamilton1992b}:
\begin{equation}
R_{H} \sim a \sqrt[3]{\frac{M_p}{3M_\star}}
\end{equation}

The capability to resolve a Hill sphere-sized gap or circumplanetary disk for a 0.1 Jupiter-mass planet at separation 1~AU, sets an angular resolution of 0.03 AU or $\sim0.2$ milli-arcseconds, requiring 10km baselines for $\lambda=10\mu$m (assuming $\theta_{\rm resolution}=\frac{\lambda}{B_{\rm max}}$).  
As the concept for PFI has evolved, we have increased the science focus on the possibility of detecting emission lines from the circumplanetary disks.  The characteristic size of this emission is likely substantially smaller than the Hill radius due to the higher temperatures needed to excite vibrational transitions in CO and H$_2$O.  Most of the 3-5$\mu$m emission from current circumplanetary disk models (e.g., Zhu\cite{zhu2015}) come from within about 3 $R_J$ which is a scale of $\sim$0.002 AU, or 0.015 milli-arcsecond at 140pc.  Note that the L-band diffraction limit for  10km baseline is $\sim0.08$~milliarcseconds, about 5$\times$ worse than the expected continuum size if heated locally.  More work is needed by the PFI-SWG to model the line emission for guiding the PFI technical requirements, with goal to measure the position-velocity diagram of orbitting gas around an accreting protoplanet.  At somewhat larger scales addressable by a 10~km PFI, outflowing gas from the disk should be detectable; much more work is needed here to understand the brightness and spatial distribution of this line emission.

For this paper, we will focus then on the possibilities of our baseline PFI design to detect 
\begin{itemize}
\item diffuse thermal emission for dust in a planet forming disk
\item the thermal emission for circumplanetary accretion disks
\item direct flux from the young planets themselves (considering ``cold start'' and ``hot start'' models).
\end{itemize}

\section{Array Design}

At this stage of the PFI planning process, the details of the telescope array geometry are not critical but a few illustrative architectures have been adopted.  In general, we have been exploring two extremes in our signal-to-noise calculations.  

 A ``minimal'' PFI would consist of 12x2.5m telescopes with 1~km baselines.  The 2.5m apertures would just allow near-IR fringe tracking on the unresolved T Tauri star fringes (see \S\ref{fringetrack}, as well as Petrov et al. in these proceedings, for more details), although 1~km would not resolve the Hill Sphere for most giant planets, except perhaps those in the outer disk.  Note that 1 km baselines would provide 2~milliarcseconds angular resolution, powerfully complementing ALMA which has almost comparable angular resolution today. In addition, we have considered a ``maximal'' PFI consisting of 21x4m telescopes.   Elsewhere in this volume (Ireland et al.\cite{ire16}), we discuss cost estimate for PFI and telescope aperture is the driving cost.  
  
To be specific, we are considering arranging the telescopes in a ring (for best UV coverage with short, spanning fringe-tracking baselines) and a Y-array (like the VLA).  The Y-array might have some practical advantages for holding vacuum beam pipes and also for re-configuring the array (if that is needed).  For Ring array, we anticipate spoke arrangement of vacuum lines to transport light from the telescopes to a central combining building, with delay lines integrated into the vacuum pipe beam transport (see discussion in this volume by Ireland et al. and Mozurkewich et al.)
 
\begin{figure} [ht]
   \begin{center}
   \begin{tabular}{c} 
   \includegraphics[width=3.5in]{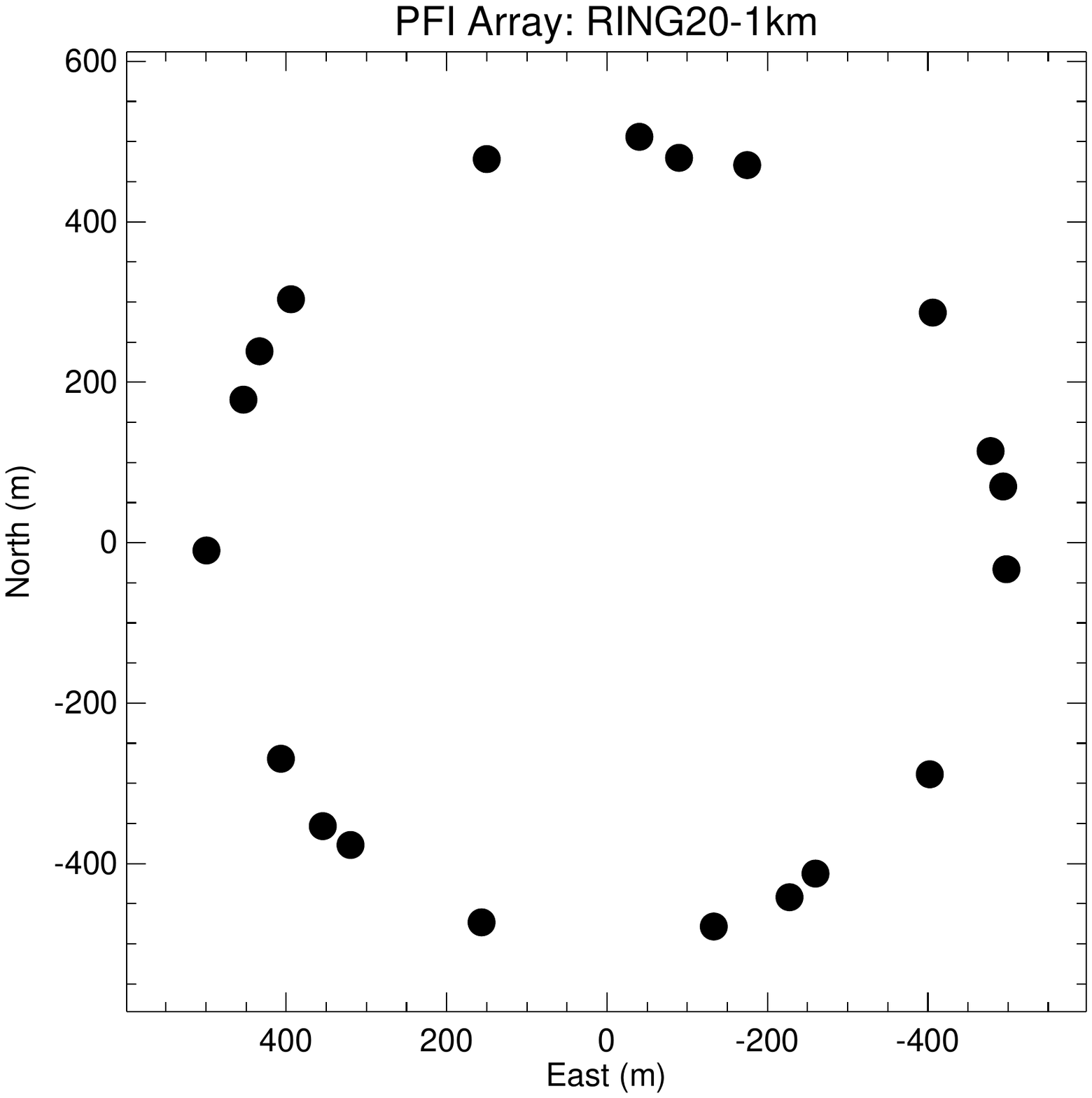}
   \hspace{-1in}
      \includegraphics[width=3.5in]{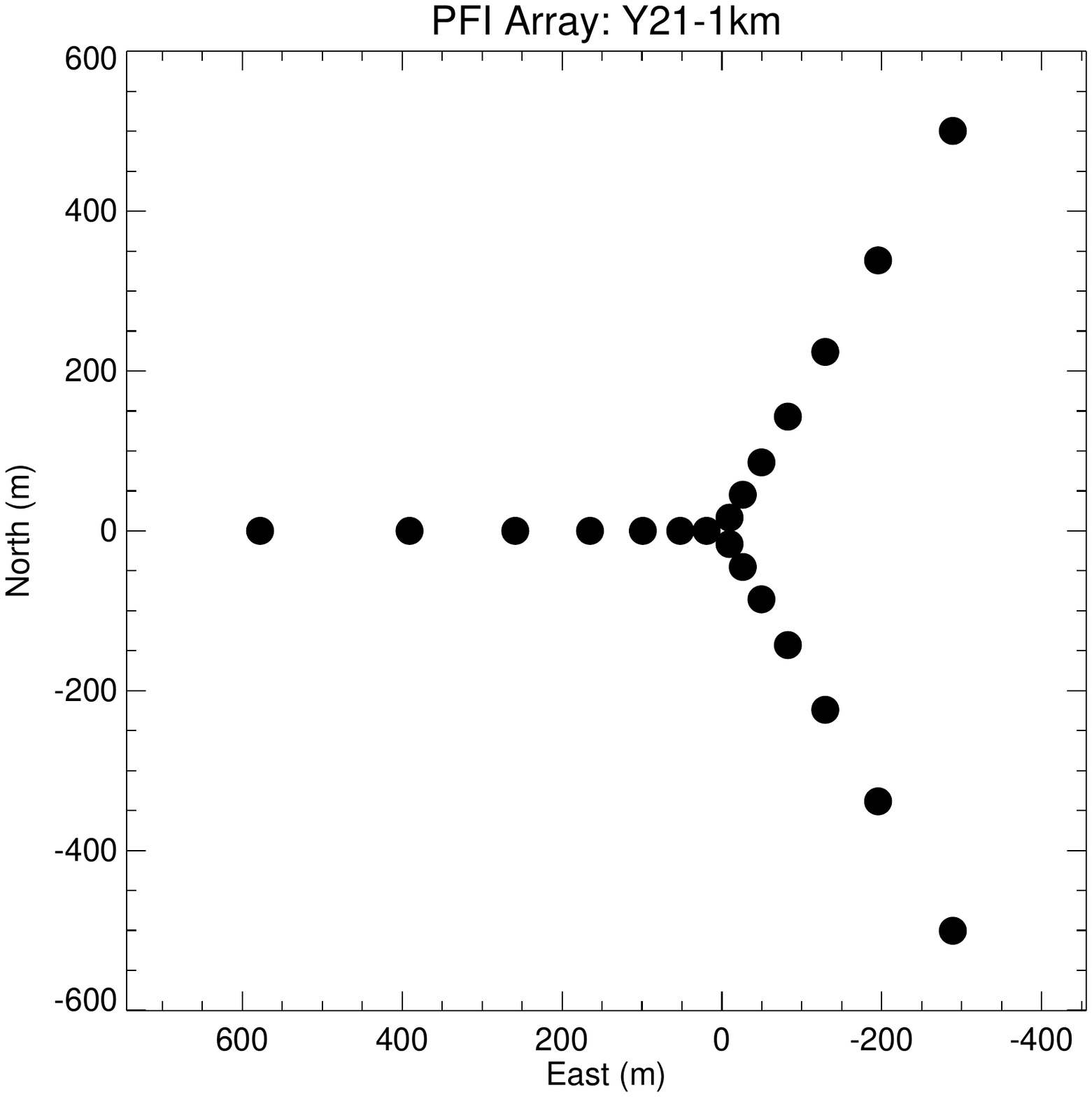}\\
      \vspace{-.4in}\\
        \includegraphics[width=3.5in]{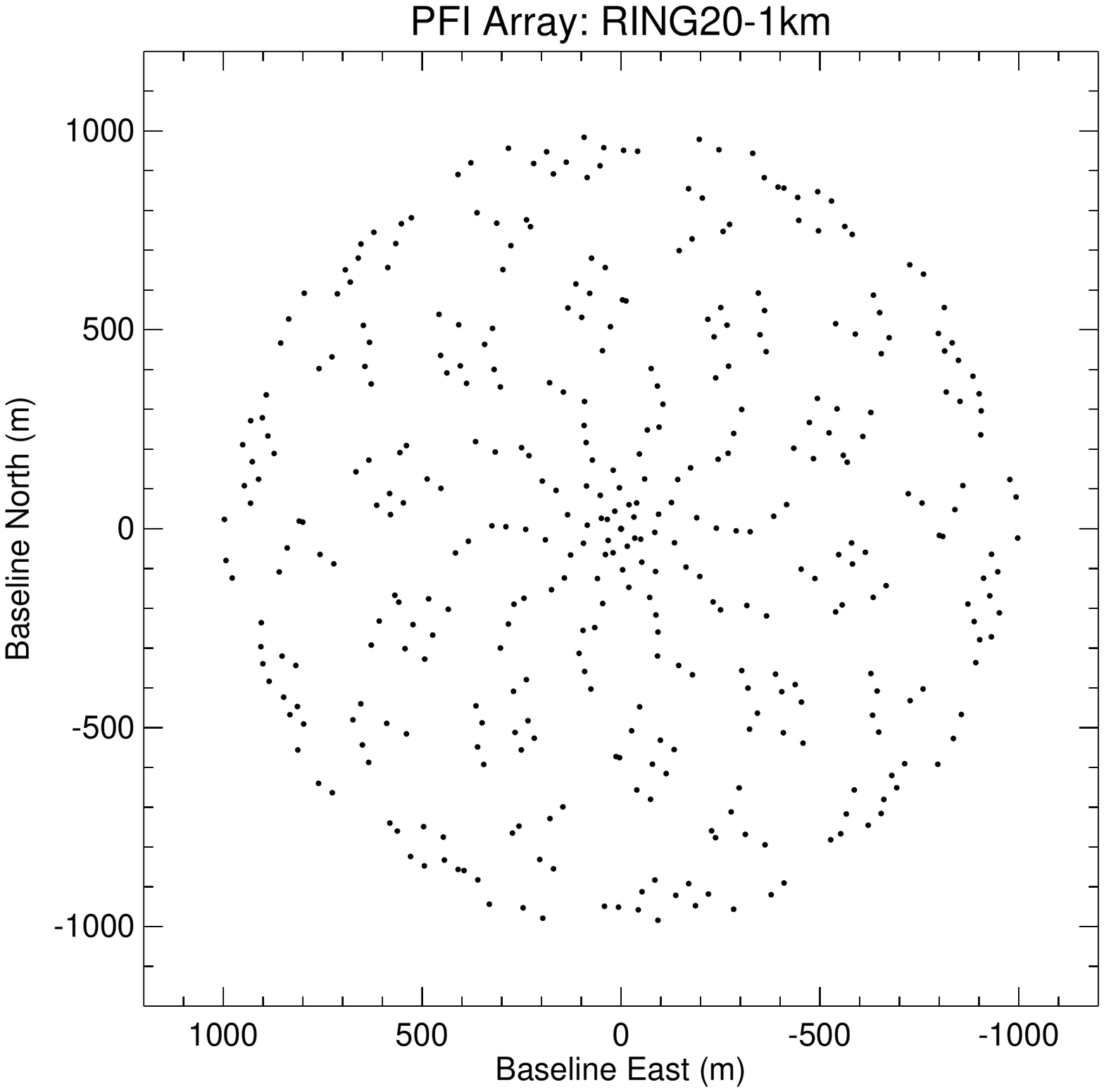}
   \hspace{-1in}
      \includegraphics[width=3.5in]{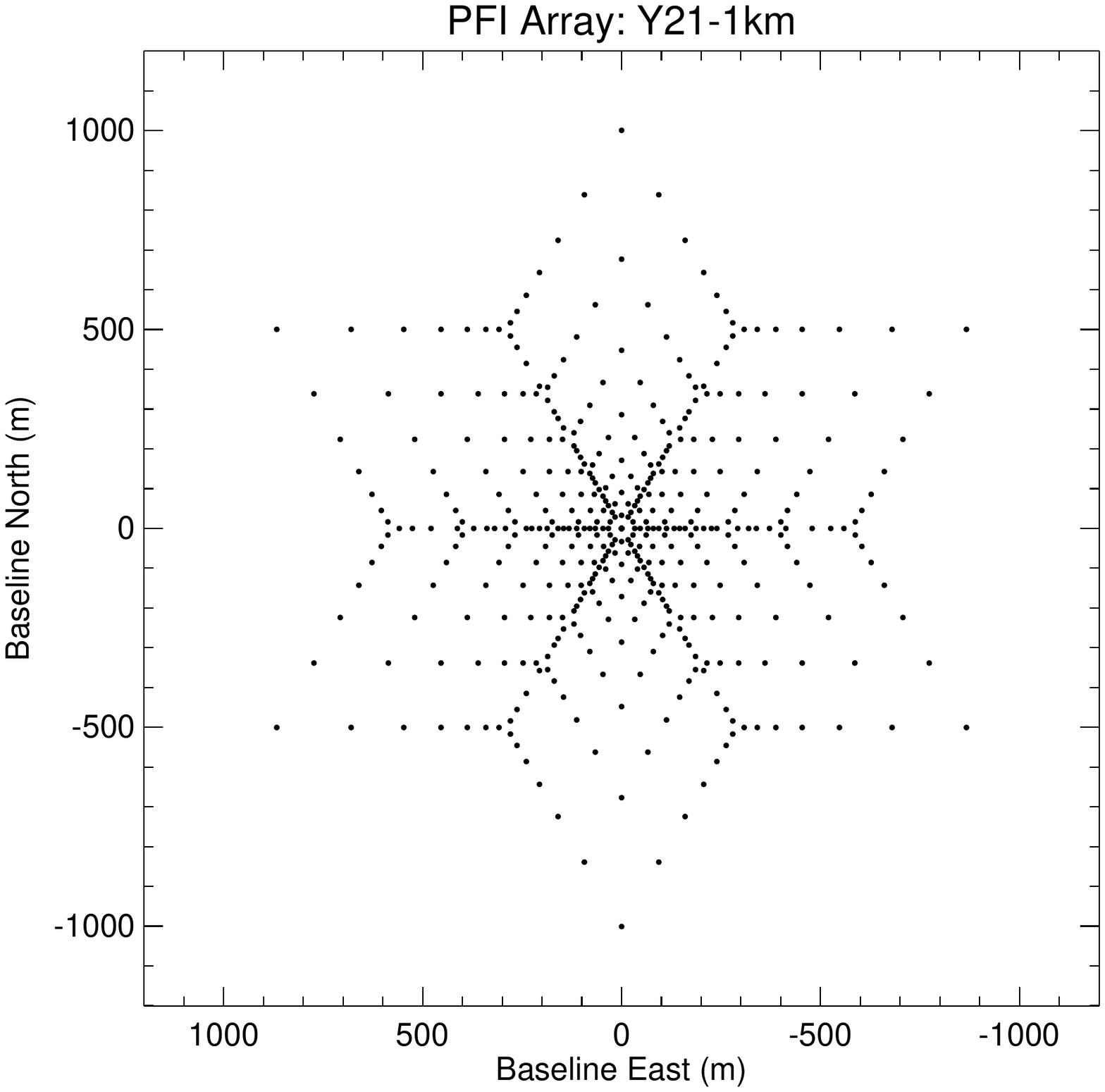}

	\end{tabular}
	\end{center}
   \caption[example] 
   { \label{fig:arrays} 
This top row shows the locations of the PFI telescopes in the Ring and Y-shaped arrays, the main geometries being considered by PFI Technical Working Group. The bottom row shows snapshot UV-coverage for the two arrays.
See Table \ref{tab:arrays} for details.    }
   \end{figure}

 Table \ref{tab:arrays} contains a summary of some array geometries that the PFI TWG are considering, while Figure~\ref{fig:arrays} show the geometry and snapshot Fourier coverages.  As expected, the ring has a more uniform snapshot coverage although with $\sim$20 telescopes, each array shown is excellent and comparable to the uv-coverage from the VLA.   Note that maximum spanning baselines are short enough so that the central star remains mostly unresolvable even for closest star forming regions.  While the primary PFI science case relies on using the unresolved emission from the central star for bootstrapping, we note that non-YSO science cases may need a more sophisticated use of closure phases and self-calibration techniques since there could be non-trivial phase signal in the bootstrapping baselines themselves.  Indeed, a NIR combiner that can supplement the fringe tracker would allow traditional closure-phase imaging for the NIR which could then be used to extract true phases for the mid-IR complex visibilities.

\begin{table}[ht]
\caption{Four Example PFI Arrays are explored in this report.} 
\label{tab:arrays}
\begin{center}       
\begin{tabular}{|l|l|l|l|l|l|}
\hline
Shorthand & Array Shape & Number of & Maximum & Minimum & Max Spanning \\
Name &  Shape & Telescopes & Baseline (m) & Baseline (m) & Baseline (m)  \\
\hline
RING20-1km   & Ring array     & 20 & 1000 &  42 & 300   \\
RING20-5km   & Ring array  & 20 & 5000 & 209 & 1500    \\
\hline
Y21-1km      & Y array & 21 & 1000 & 33 & 187 \\
Y21-5km      & Y array & 21 & 5000 & 165 & 935   \\
\hline

\end{tabular}
\end{center}
\end{table}

At this preliminary stage of PFI trade studies, we focus on a sensitivity analysis, assuming that the ring and Y-array both can achieve excellent Fourier coverage.  We expect practical issues will decide which geometry is eventually recommended. In this report we will only show imaging simulations results for the Y-array with 21x2.5m telescopes.  SNR scales like $D^2$ and $\sqrt{n_{tel}}$ (direct detection) or $\propto n_{tel}$ (heterodyne).

\section{Beam Transport}

The TWG sub-committee on beam transport has a contribution in these SPIE proceedings (Mozurkewich et al). Here we give an overview of a few methods for beam transport that are being considered:
\begin{itemize}
\item Single-mode fiber optics.  The advantages in cost are traded against versatility since only high transmission fiber exists in the near-infrared.  A fiber-only beam transport is only sensible for a limited PFI that uses J/H band fringe tracking with mid-IR heterodyne, since the latter method does not require any direct beam transport since the light is downconverted at each telescope. As mentioned above, with the emerging excitement about L/M band operation, we tend to disfavor this idea.
\item Cooled, vacuum pipes.  While a simplistic beam transport of a collimated beam would require relatively large optics, a sophisticated beam transport system that involves dynamically transporting the pupil while sending light beam through a focus can allow a smaller diameter beam train (see contributions by  Ireland et al, Mozurkewich et al., in this SPIE).  Different wavelengths might require different beam trains, depending on optical solution.  The PFI-TWG is confident a solution exists here with existing technology, although new optical design work is needed. One complication is that the optics will need cooled to minimize thermal emission, the dominant noise source for PFI.
\item With the advent of inexpensive adaptive optics, one could consider an optical link for beam transport. This method would mean transporting the telescope light over open air through line-of-sight mirrors.  A bright LED is used as a guide star and a deformable mirror will correct the light, either pre-shaping or  post-transport.  Note we also need a path length monitor to track piston variations.  For very long baselines, the differential air path might limit the coherence length but could be solvable through a longitudinal dispersion compensation system (using prisms of different glasses) or using the ring geometry which has near equal distances from the ring center.  It is unclear if this system might make accurate wavelength boot-strapping more difficult, although a series of near-IR LEDs could monitor ground-layer air dispersion quite accurately.  Note that the ISI interferometer already has achieved open-air beam transport of their local oscillator laser, without AO but including active path length monitoring and compensation (Hale et al.\cite{hale2000}). Due to extra reflections needed, expect this sytem to have lower overall throughput than the cooled, vacuum pipe solution. 
\end{itemize}
In conclusion, the PFI-TWG has identified workable beam transport options; see Ireland et al. and Mozurkewich et al (these proceedings) for more information.

\section{Interferometer Model}

\subsection{General considerations}
A fundamental limit for PFI is that fringes will never be detected in a single coherence time in the  mid-infrared for T Tauri stars. This necessitates the need for a near-infrared fringe tracker that will allow coherent integration of the mid-infrared interference patterns.   Furthermore,  individual complex visibilities will have low SNR, thus image reconstruction algorithms need to rely on complex visibilities and not visibility-squared, the closure phases, or bispectral amplitudes that all suffer much worse SNR.

\subsection{Fringe tracking considerations}
\label{fringetrack}
Petrov et al and Minardi et al (these proceedings) discuss fringe tracking and beam combination for PFI in more detail. Here let me simply remind the reader that with 10\% transmission, a 2.5m telescope will collect $\sim$11000 H-band photons for a T Tauri star in Taurus ($H=8.1$) in a 10-millisecond exposure, a time short enough to ``freeze'' the atmospheric fluctuations in the near/mid-IR.  If we further divide this into 8 spectral channels (needed to separately model the dry air and moist air components) and we share light 4 ways  to interfere with at least 4 other baselines, we are then down to a best-case of $\sim$340 photons per telescope-baseline.  For a noiseless detector, this should be sufficient raw SNR for fringe tracking with phase rms $<$ 0.2 radian for baseline bootstrapping up to 1km baselines from nearest-neighbor fringe tracking.  Needless to say, no current fringe tracker is so parsimonious with photons but PFI will need to make high throughput an absolute priority.

While 2.5m telescopes are adequate for near-IR fringe tracking, we noted already that SNR for background limited observing is proportional to telescope area.  Another way to say this, is that for fixed SNR, the required integration time scales like $t_{\rm int}\propto D^{-4}$, thus a 4m has 6.5-fold time advantage over a 2.5m telescope for PFI.

One might think that a major risk for PFI would be the feasibility to use near-infrared fringes to accurately and robustly phase up mid-infrared light. This is not trivial since the index of refraction of air in the mid-IR is dominated by water vapor while dry air dominates the visible and most of near-infrared.  That said, this problem has already been tackled expertly by the 85-meter baseline Keck Interferometer for NASA's Nuller project.  Colavita\cite{colavita2010} demonstrated that the measurement of slightly dispersed fringes in K band was sufficient to account for the varying amount of dry and moist air and to feed-forward/predict the optical path difference (OPD) for the mid-infrared.  They demonstrated coherence for over 5 minutes which should be adequate time to use a bright nearby calibrator for fine-tuning the group/phase delay adjustments. Colavita used analytic results to show that H-band data alone would also be sufficient for this purpose. Further, work on MIDI\cite{pott2012} has also supported this conclusion and future MATISSE studies will use a K band GRAVITY Fringe tracker to simultaneously observe at L, M, and N bands. We even heard a new report on successful feed-forward fringe tracking at the 2016 SPIE meeting for the LBTI by Defr\`{e}re et al.

\subsection{SNR equations}

While the fringe tracking combiner architecture is non-trivial to design (since not all baselines are needed), we insist on measuring all mid-IR baselines given the complexity of the imaging targets.  We believe a pair-wise system is not practical and adopt an  ``all-in-one'' combiner for the PFI science instrument (see Ireland et al. and Minardi et al. in these proceedings for a specific implementation).  In this combiner, each fringe measurement is contaminated by background photons from ALL the beams, thus suffering a $\sqrt{n}$ noise penalty for $n$ telescopes.  

While there are many small assumptions that affect the exact form of the SNR equation, we adopt a simple formulation for complex visibility signal-to noise ratio S/N on each baseline: 
\begin{equation}
(\frac{S}{N})_{\rm direct} = \frac{ \eta_T F_{0,\lambda} 10^{-{\rm mag}/2.5} V_{\rm baseline} (\frac{\pi}{4}D^2) \sqrt{2 t_{coh} \Delta\lambda  }}  {\sqrt{  \epsilon_{\rm warm} N_{\rm tel} B_\lambda(T) \lambda hc  } }
\end{equation}
where $\eta_T$ is the total throughput of system, $F_{0,\lambda}$ is the flux zero point, mag is the target flux magnitude, $V_{\rm baseline}$ is the visibility amplitude on that baseline,  $D$ is telescope diameter, $t_{coh}$ is the coherent integration time, $\Delta\lambda$ is the bandwidth, $\nu$ is observing frequency, $\lambda$ is observing wavelength, $\epsilon_{\rm warm}$ is the product of the cold throughput of system times the emissivity of the warm optics at temperature T, $N_{\rm tel}$ is the number of telescopes, $B_\lambda$(T)  is the Planck function for temperature T.  All the adopted parameters are included in Table~\ref{tab:snr}.

Similarly for heterodyne detection (see discussion also in these proceedings by Vasisht et al. and Michael et al.), we assume that shot noise from local oscillator (or alternatively vacuum noise from the target field) is the only source of noise and we obtain:

\begin{equation}
(\frac{S}{N})_{\rm heterodyne} =  \frac{ \eta_T F_{0,\lambda} 10^{-{\rm mag}/2.5} V_{\rm baseline} (\frac{\pi}{4}D^2)  \lambda^2 \sqrt{2 t_{coh} \Delta\lambda}  }{ hc^{3/2}  }
\end{equation}
where for heterodyne we assume a higher throughput than direct detection case (see Table~\ref{tab:snr}).

Note that for the assumptions in Table~\ref{tab:snr} that the heterodyne and direct detection schemes have equal SNRs at 10 microns for $N_{\rm tel}\sim$40 but this estimate changes dramatically with assumptions about the throughput and beamtrain warm emissivity. For L-band, heterodyne is never as sensitive as direct detection, even with thousands of telescopes.  However, heterodyne should be more sensitive for the interesting atmospheric windows at 24$\mu$m and 40$\mu$m, if laser comb local oscillators and high speed detector arrays could be developed (see Ireland \& Monnier\cite{ireland2014spie} for more details on scheme).

\begin{table}[ht]
\caption{Basic assumptions of image reconstruction presented here} 
\label{tab:snr}
\begin{center}       
\begin{tabular}{|l|c|}
\hline
Parameter & Value \\
\hline
Target & TTS in Taurus\\
Telescope Diameter (m) &2.5 \\
Array Shape & Y-array \\
Max Baseline (km) & 1 \\
Number of Telescopes & 21 \\
FOV (arcsec)   & 0.5 \\
Coherent $t_{\rm int}$ (min) & 1.0  \\
Wavelength Bandpass ($\mu$m) &10--11 \\
$\frac{\lambda}{\Delta\lambda}$ & 250  \\
Time On-source/night (hrs) & 10 \\
Number of Observing Nights & 3 \\
\hline
\hline
\multicolumn{2}{|l|}{Direct Detection Combiner}\\
\hline
Mid-IR Throughput & 0.17 \\
Warm Optics Emissivity & 0.27 \\
\hline
\multicolumn{2}{|l|}{Heterodyne Combiner}\\
\hline
Mid-IR Throughput & 0.35 \\
\hline
\end{tabular}
\end{center}
\end{table}

\section{Image Reconstructions}
\subsection{Observing simulation}
   
While finding a suitable dry site for a 10km interferometer with excellent seeing is non-trivial (see discussions in Ireland et al., these proceedings), we must adopt a location for our simulations here.  We place PFI at mid-latitude site in the Northern Hemisphere, specifically on Mt. Wilson, CA.  We make our calculations for observing a source located in Taurus  on UT November 25, 2030 -- our model will be the 4-planet transition disk of Dong et al\cite{dong2015}.

For PFI, data can be combined from many nights to uniformly sample the Fourier plane but for here we will simply assume we get complete data from 7pm to 5am local (3:00 to 13:00 UT time), essentially the entire hour angle range with the object above 22 degrees elevation.  While in actual observations there would be gaps in time as we observe calibrator targets, we will ignore this and assume we can remove all gaps through re-visiting targets over many nights.  This is a total on-source integration time of 10 hours per night.

A number of issues can limit the field-of-view (FOV) of our reconstructed images, For our simulations we plan to maintain a 0.5'' FOV.  In order to minimize field-of-view limitations due to time-averaging effects, we will average data on 1~minute time scales for 1km array (see Monnier \& Allen\cite{monnier_psss}).  Bandwidth-smearing is another serious effect and to maintain a 0.5'' FOV we need to observe with spectral resolution $R=\frac{\lambda}{\Delta\lambda}=250$ for maximum baseline 1km. We have used the entire 10--11$\mu$m wavelength range for the image reconstructions, assuming grey emission.

As discussed earlier, the telescope aperture is most critically set by the ability to fringe track effectively and for this section we will assume $D=2.5$m to be concrete.

   \begin{figure} [ht]
   \begin{center}
   \begin{tabular}{c} 
   \hspace{-.2in}
   \includegraphics[width=3in]{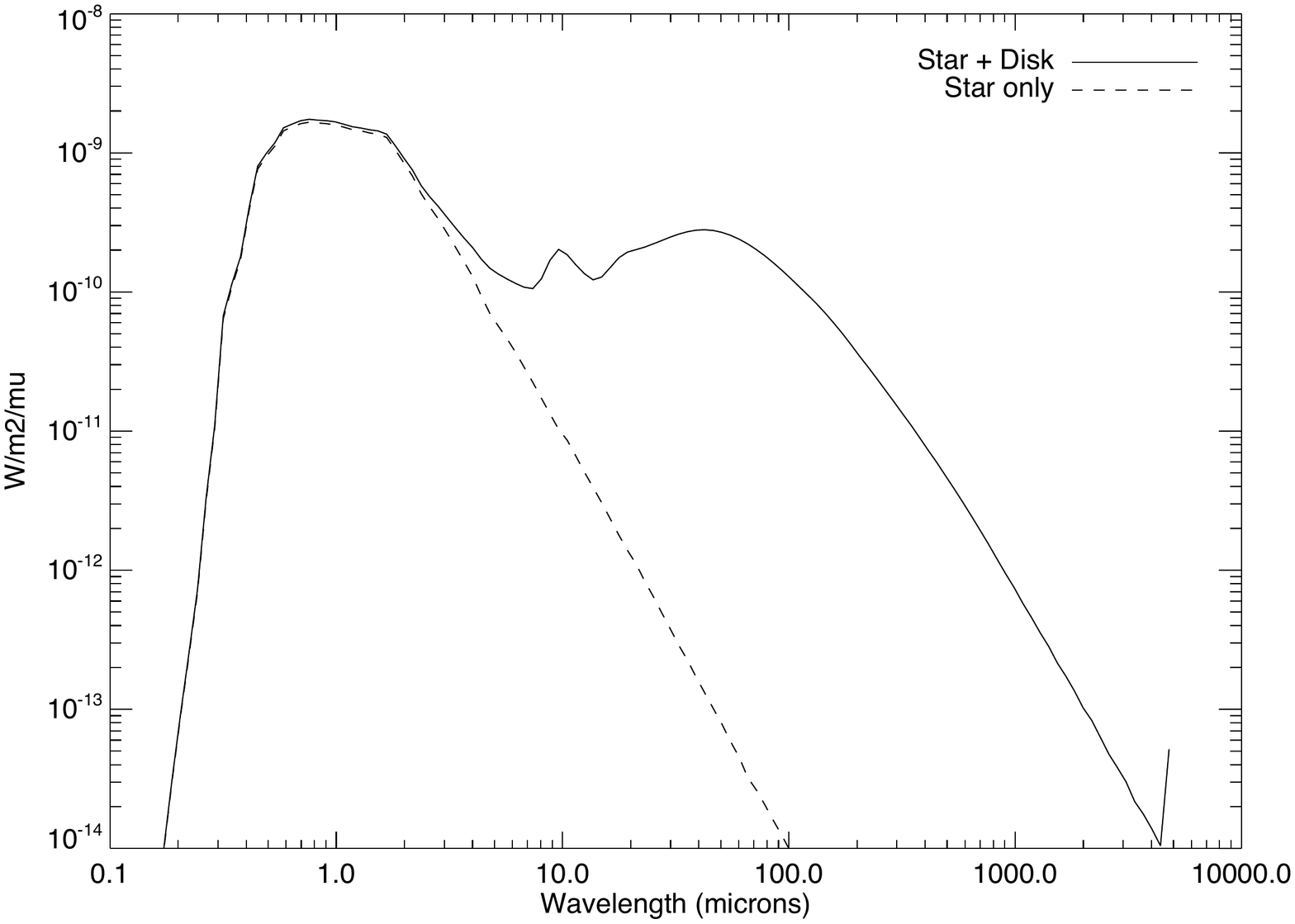}
   \hspace{-.5in}
      \includegraphics[width=3in]{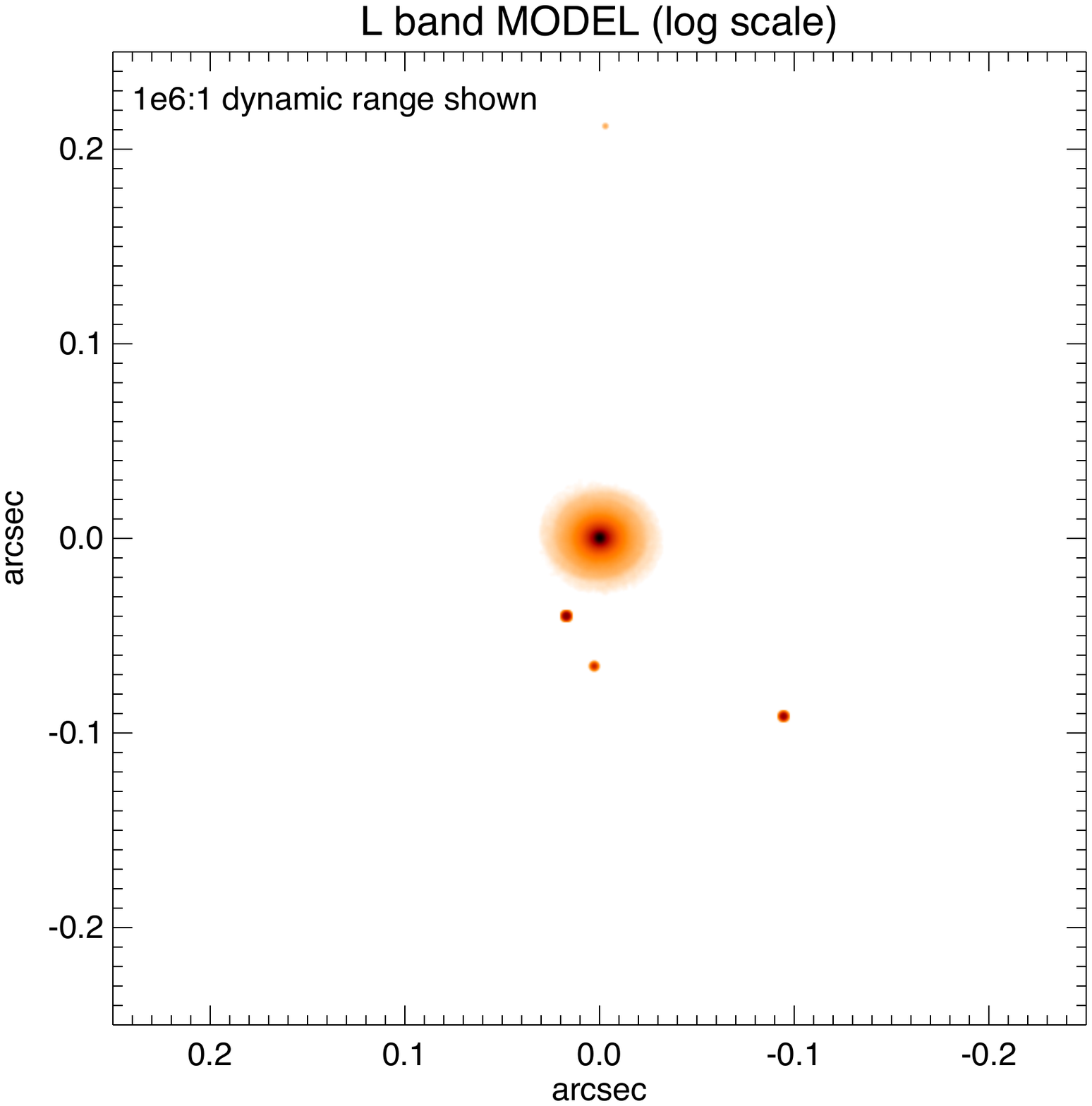}
      \hspace{-1in}
            \includegraphics[width=3in]{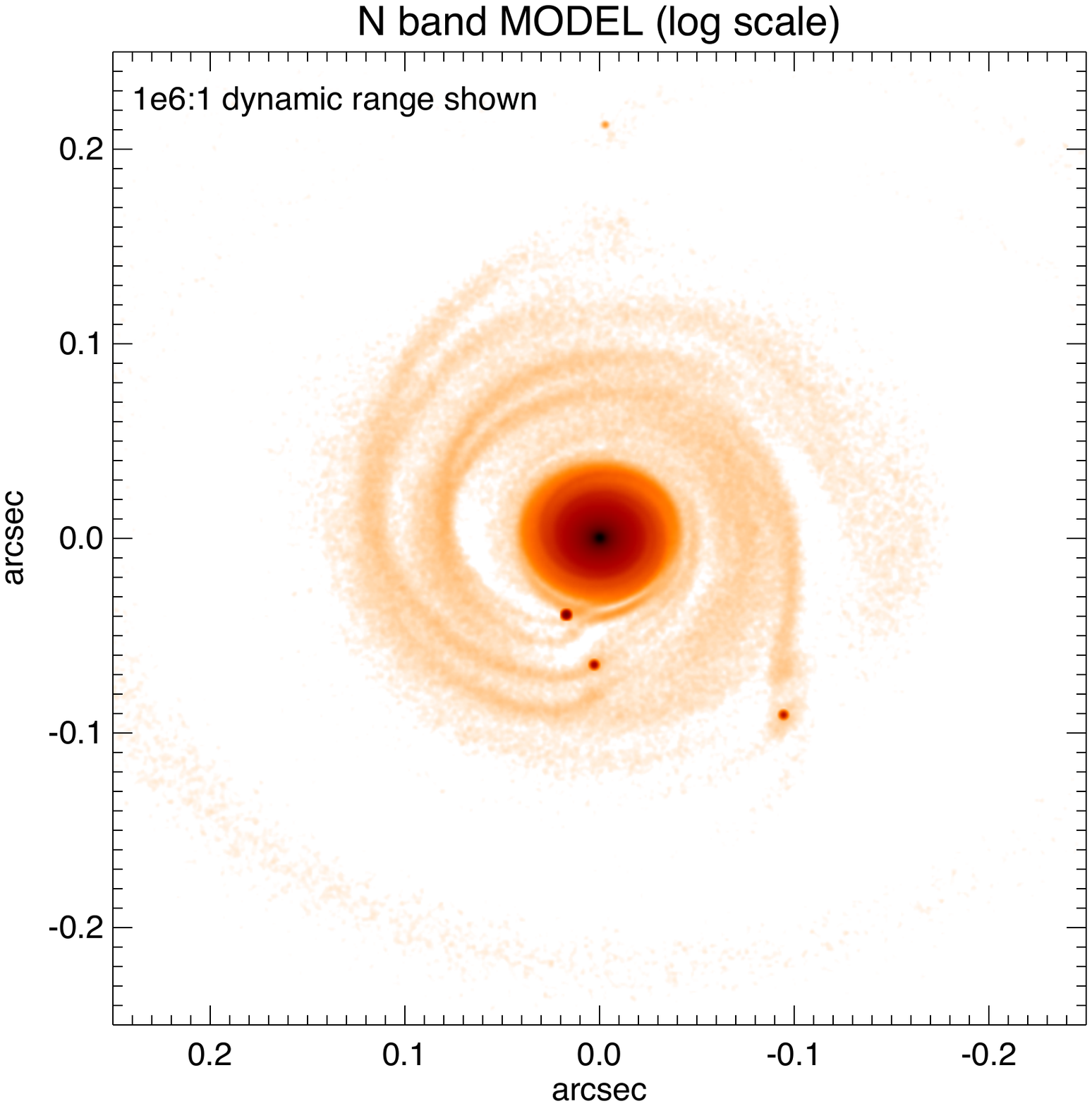}
	\end{tabular}
	\end{center}
   \caption[example] 
   { \label{fig:model} 
We used a 4-planet gapped disk from Dong et al\cite{dong2015} for creating a test data set for
PFI simulation. Here we see the input model including a) SED, b) L band image, and c) N band image.  In these images, we show the images with log scaling with a $10^6$:$1$ dynamic range.  Here, the star and 4 protoplanets are smoothed to 2.0 mas resolution, roughly the diffraction limit for 1 km PFI in N band.}
   \end{figure} 

\subsection{Models used for synthetic data}

  We used a 4-planet gapped disk from Dong et al\cite{dong2015} for creating a test data set for
PFI simulation.  We present the SED and radiative transfer images in Figure~\ref{fig:model}.   These models did not include self-luminosity from the planets themselves nor from circumplanetary accretion disks. We have added these emission sources into our synthetic images using estimations from Zhu\cite{zhu2015} -- these assumptions are found in Table\,\ref{tab:flux}.
Specifically we will use $M \dot{M} = 10^{-6}$ to $10^{-5} M_J^2$ yr$^{-1}$ and $R_{\rm inner }= 1.5 R_J$. The circumplanetary accretion disk at N band absolute magnitude (d$=10$~pc) ranges from mag $6.5$--$3.7$ while the exoplanet itself ranges from $7.5$--$12$ depending on the mass and hot/cold start assumption.  The four planets were formed at 7.5 AU/11.9 AU/18.9 AU/30.0 AU where we have assumed the inner two are dominated by accretion luminosity from their disks while the outer two planets are hot start/cold start 10~$M_{\rm Jup}$ models with no disk emission.  As discussed later, we will want to explore additional planet formation scenarios in the future as the diffuse disk emission beyond 5~AU is difficult to detect at 10$\mu$m for most ground-based PFI arrays we have considered.


\begin{table}[ht]
\caption{Assumptions used in simulating PFI imaging performance (absolute magnitudes reported below for assumed distance 10\,pc; add distance modulus 5.7 magnitudes to locate at 140pc)} 
\label{tab:flux}
\begin{center}       
\begin{tabular}{|l|c|c|c|c|l|}
\hline
Component & $M_R$ & $M_H$ & $M_L$ &$M_N$ & Reference\\
          & (AO & (fringe & (dust & (dust &\\
          & system)  & tracking) & \& planets) &  \& planets)&\\
\hline
Example T Tauri Star & & & & & Baraffe et al.\cite{baraffe1998} (1998) \\
\qquad 1 Msun, 2.1Rsun, 3865K& 5.87 & 2.54 & $\sim$2.5  & $\sim$2.4 &\\
\qquad 3 Myr, [Fe/H]=0.0 &&&&&\\
\hline
Protoplanet & & & & & Spiegel \& Burrows\cite{spiegel2012}  \\
\qquad ``hot start" 10MJ & &8.7 & 7.8 & 7.5 & (2012) \\
\qquad ``cold start" 10MJ & & 16.8 & 14.1 & 11.9 &  \\
\hline
Circumplanetary Disk & & & &&   Zhu\cite{zhu2015} (2015) \\
\qquad ($R_{in}=1.5~M_J$) & & & && \\
\qquad  $ M\dot{M}=10^{-5}M^2_J\,{\rm yr}^{-1}$ & & 10.6 & 6.9 & 4.6 & \\
\qquad $ M\dot{M}=10^{-6}M^2_J\,{\rm yr}^{-1}$ & & 16.4 & 9.8 & 6.5 & \\
\hline
4-planet gapped disk & & & && Dong et al\cite{dong2015} (2015) \\
\qquad Star only (2Rsun, 4500K) & 4.8 & 2.1 & 2.1  & 2.1 & \\
\qquad Star + Disk (30$^\circ$ inclination)                & 4.8 & 2.1 &  1.6 & -1.1 & \\
\hline
\end{tabular}
\end{center}
\end{table}

\subsection{Results using CLEAN}
As discussed in the last section, we find ourselves in the imaging regime of ALMA not CHARA. This means that much of the work on imaging algorithms over the past decade in optical interferometry through SPIE beauty contests, e.g., BSMEM, MIRA, MACIM, is not directly applicable to PFI image reconstructions.  Indeed, these generally relied on the $V^2$ and bispectrum and involved heavy use of regularizers that make a PFI reconstruction numerical intractable with current approaches.  Baron et al. (elsewhere in these proceedings) will discuss novel ways to solve the numerical problems, but for now we are left to adopt a simple CLEAN algorithm for our first simulations.

   \begin{figure} [ht]
   \begin{center}
   \begin{tabular}{c} 
   \hspace{-.2in}
      \includegraphics[width=4in]{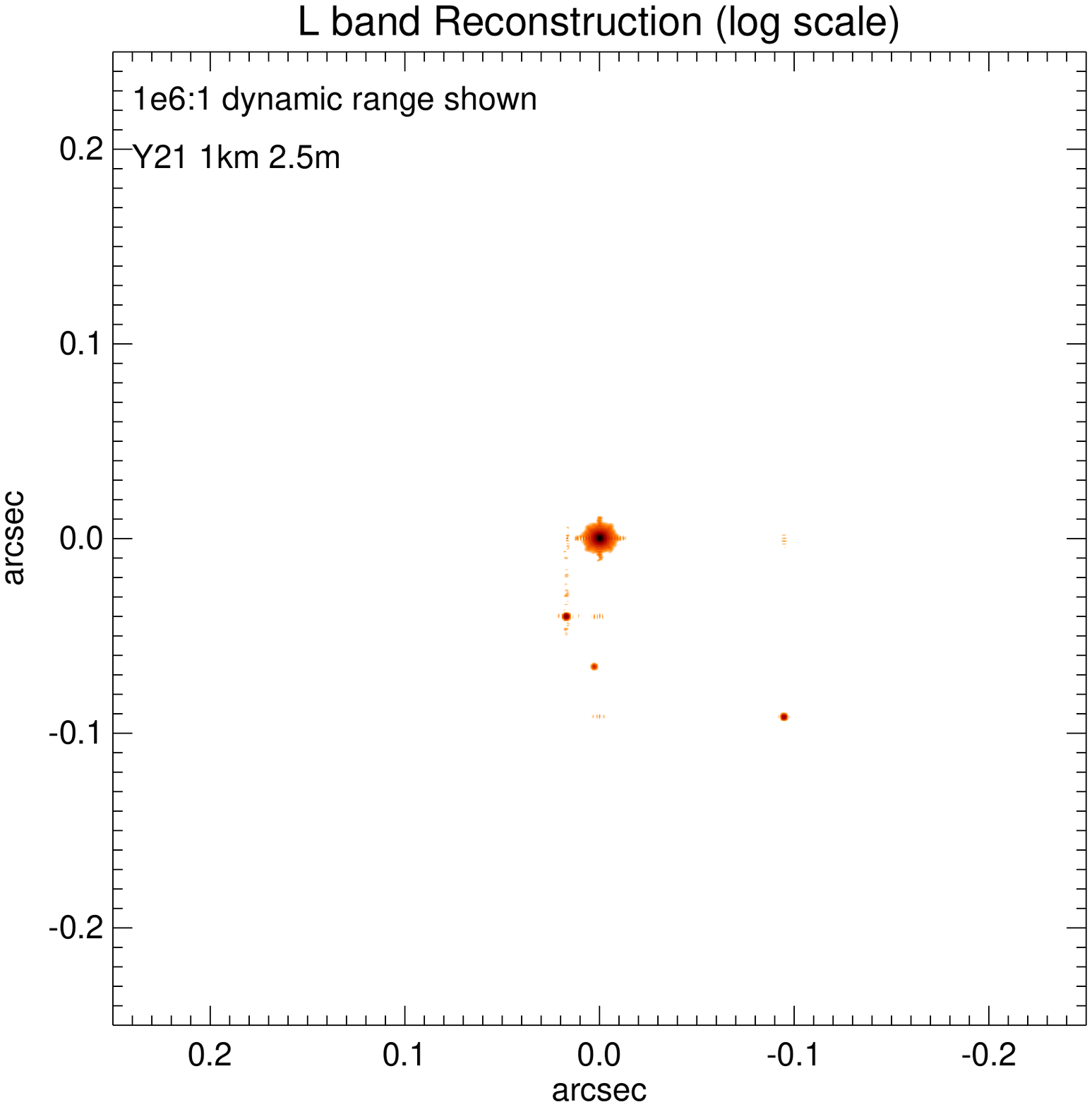}
      \hspace{-1in}
            \includegraphics[width=4in]{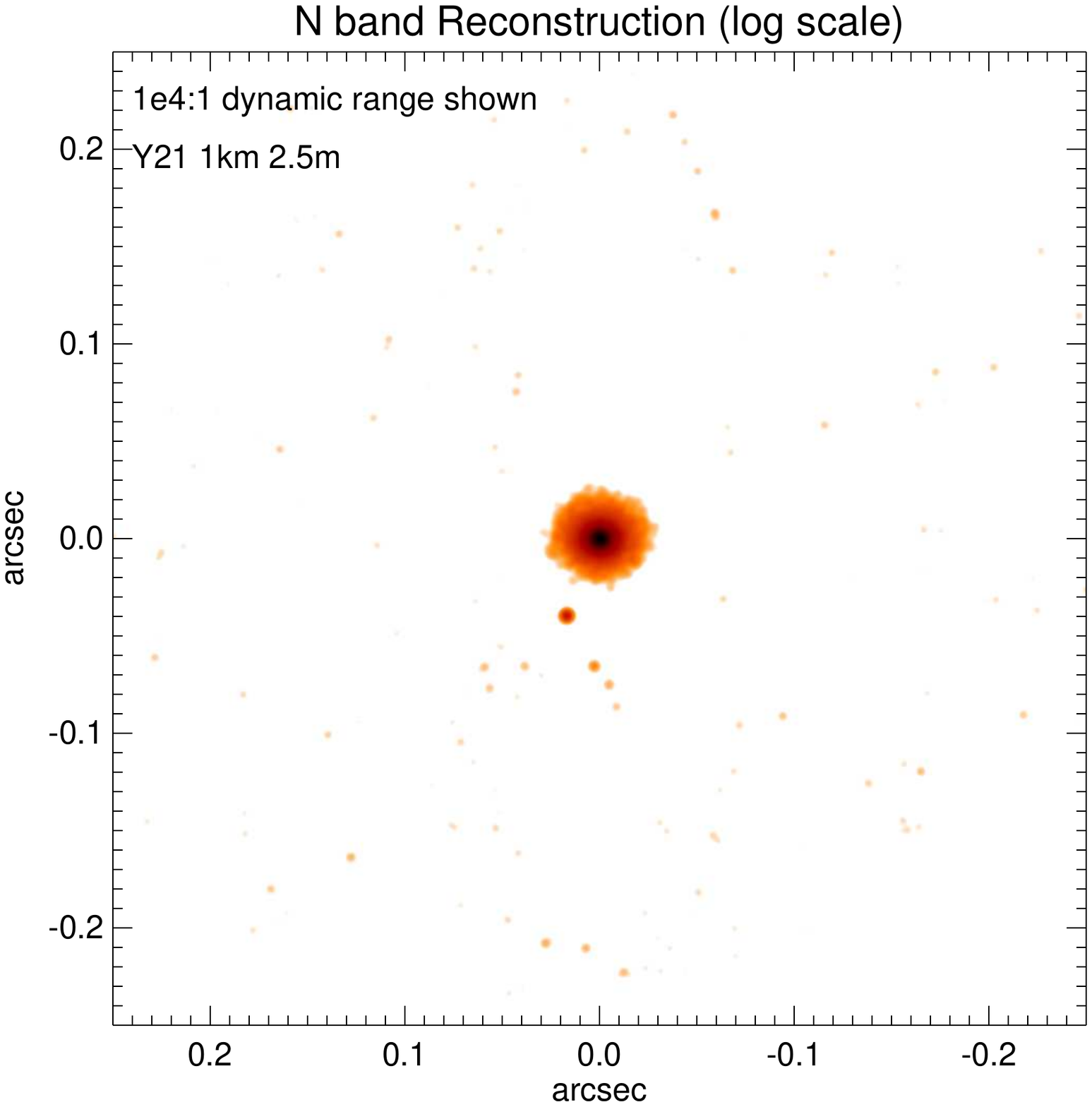}
	\end{tabular}
	\end{center}
   \caption[example] 
   { \label{fig:results} 
Results from our image reconstructions using a simple CLEAN algorithm.  We detect the accretion disks around both the inner planets in L and N band, but the planets themselves are picked up mainly at L band.}
   \end{figure} 

Figure~\ref{fig:results} shows our image reconstructions.  We compared the point source sensitivity and surface brightness limits here with the analytic derivations found in Ireland et al. (these proceedings).  Indeed, our results confirm a L-band point source sensitivity of about mag 19 and N-band limit of 13, while finding a N-band surface brightness of $\sim$150K matches the results seen in Figure~\ref{fig:results}. 

We note that the exciting gaps and structures seen beyond 5 AU in the the synthetic images required $1:10^6$ dynamic range while we our reconstructions only obtained about $1:10^4$.  We believe that other simulations would present more favorable scenarios for PFI, since this example had all the planets in the outer disk and dust in the inner disk blocked the outer disk material, drastically lowering the surface brightness of the most interesting features.

On the other hand, the simulations did show the immense potential for PFI to measure the spectral energy distributions (SEDs) of accreting protoplanets from L-N band, an exciting prospect indeed.  One must ask whether an ELT coronagraph will be able to do this more easily than an interferometer, since an ELT must extract this signal within a few $\lambda/D$ and with presence of confusing dust emission.  Even a "short baseline" 1km PFI will easily resolve out dust structures on the long baselines and easily pick out the emission from accreting Jupiter-mass planets.

\subsection{Notes on imaging simulations}
Based on this imaging experiment as well as those explored by Baron et al. elsewhere in these proceedings, we summarize some key points here:
\begin{itemize}
\item The original code to generate synthetic data broke in both the IDL and C versions. This was partially solved by only generating complex visibilities since visibility-squared and bispectral quantities were too noisy anyway to be used.
\item PFI datasets are similar in scale to ALMA. This caused OI-FITS readers/writers in IDL to break due to the large size of FITS files 
\item Image reconstructions were done using only complex visibilities, assuming perfect fringe tracking.  This is essentially identical to conventional radio/mm-wave interferometry
\item None of the imaging algorithms developed for optical interferometry and tested during past SPIE ``Beauty Contests'' worked well at first.  Some failed because of the large datasets causing memory or integer overflows, others failed because they were not written to use solely complex visibilities, while most simply were too slow to converge in a human lifetime.  Most of the algorithms can be salvaged with special attention to speed issues, but this will require focused effort.
\item All reconstructions here were done with simple home-made CLEAN algorithm in IDL.  Note that LEAN-based algorithms do not handle multiple spatial scales well.
\item The synthetic PFI data was imported into CASA for testing but preliminary results from using CASA's version of  clean and multi-scale clean were not superior to IDL implementation.
\item Effort should be put into fast image reconstructions using complex visibilities and regularizers.  This will likely require efficient GPU parallelization and could find useful application to ALMA data (see Baron et al).
\end{itemize}

\section{Preliminary Conclusions}
While we have only just begin to seriously confront models and image reconstructions with realistic PFI array geometries, we present a list of preliminary conclusions from the work of the PFI Technical Working Group:
\begin{itemize}
\item We have validated both numerical and analytic estimates of the PFI sensitivity for our baselines architectures.
\item Image reconstructions of synthetic data based on realistic planet formation simulations show that PFI can filter out diffuse emission to focus in on young planets and their accretion disks, immune to confusion that will be a limiting factor for imaging with ELTs.
\item Future PFI work will  explore imaging young planets and their accretion disks.  Can we resolve the continuum and/or line emitting regions? We currently lack realistic  models.
\item A ground-based mid-infrared version of PFI will have  N-band surface brightness limited to $>$100K, or even 150K. PFI Science Working Group and TWG will coordinate to explore what stages of planet formation can be imaged with our PFI baseline architecture.  For instance, cooler outer dust will be much more visible if directly heated by stellar emission, not blocked by an optically-thick inner disk. Currently we avoid strong conclusions on what kinds of diffuse dust emission features can be probed until a wider range of disk simulations are considered -- we want to guard against over-interpreting the results of a single simulation.
\item Class 0 objects may have strong local heating sources from exoplanets and accretion -- this needs to be explored, but models are not available.
\item PFI TWG is focusing on ways to increase sensitivity  and wavelength coverage -- currently UV coverage is not a limiting factor on the science retrieval.
\item Sensitivity and wide-wavelength coverage pushes design toward a PFI with fewer, larger telescopes with vacuum beam transport rather than many small telescopes. This is not friendly to the advantages of heterodyne, although final SNR trade-off conclusion depends sensitively on the assumed warm emissivity and cold throughput.  
\item PFI SWG/TWG will investigate science potential and technology readiness for 24$\mu$m and 40$\mu$m heterodyne observations that could be possible from dry sites.  In one scenario, a far-IR heterodyne channel would be an exciting "add-on" to a direct detection PFI.
\end{itemize}

\section{Technology Roadmap}

The TWG has developed a series of key technologies that would greatly benefit the affordability and technical feasibility of PFI.  We give an overview of key areas and present them in Table~\ref{tab:roadmap}

\begin{table}[ht]
\caption{PFI Technology Roadmap} 
\label{tab:roadmap}
\begin{center}       
\begin{tabular}{|l|l|}
\hline
Critical Technology &  Considerations \\
\hline
Inexpensive Large Aperture Telescopes & 1) Replicated parabolic lightweight mirrors\\ 
& e.g., Carbon-fiber reinforced polymer (CFRP)\\
& 2) Spherical primary + Gregorian secondary corrector \\
 & e.g., either CFRP spherical primary or \\
 & glass segments on CFRP truss\\
 & {\bf Partner with industry, engineers, DARPA}\\
\hline

L/M band IO combiners & Needed for high precision calibration \\
& {\bf Explore Chalcogenide integrated optics }\\
\hline
Wavelength-bootstrapped fringe tracking & L band imaging require $10^6$:$1$ dynamic range imaging \\& ultra-accurate fringe tracking in L based on H-band\\
& {\bf Pilot project at MROI or VLTI}\\
\hline
Mid-IR Laser Comb Heterodyne & Possible "add-on" to L/M band ``direction detection" array\\
 & Explore 24$\mu$m, 40$\mu$m atmospheric windows \\
 & {\bf Develop toolkit: combs, detectors, digital processing} \\
\hline
Image reconstruction software & Optimally integrate data from multiple scales,\\
& e.g., 30m telescope imaging $+$ PFI interferometry, \\
& Ultra-high dynamic range imaging exoplanets detection\\
& {\bf Synergize with modern algorithms for ALMA }\\
\hline
Space interferometry	& Longer-term future for high sensitivity. \\
 & Demonstrate formation flying interferometry with cubesats\\
 & {\bf Support new missions, see Rinehart et al. }\\
\hline
\end{tabular}
\end{center}
\end{table}

\subsection{Telescopes}
Even with twenty-one 2.5-m telescopes, our simulations show that PFI has limited sensitivity to the diffuse low surface brightness dust emission and also to faint emission from all but the most vigorously accreting young planets.  For the most part, the signal-to-ratio for PFI scales like telescope area while only increasing like $\sqrt{N_{\rm telescopes}}$, thus putting a premium on large telescopes for PFI.  
Ireland et al. showed that telescopes are the dominant cost when unit telescopes are more than $\sim$\$4M each, thus any progress in telescopes manufacturing methods will have a profound affect on the science potential of PFI at a given price point. Given the engineering challenges of building intermediate-sized telescopes and enclosures, progress here will require creating new  partnerships with industry.

\subsection{L/M band integrated optics combiners}
A number of presentations at SPIE explored new developments in this active area, see Labadie et al, Minardi, et al, Kenchington-Goldsmith et al.  In addition to fundamental explorations of new materials and methods, there are opportunities for exciting on-sky tests at MROI, CHARA or VLTI in the PFI-related science of YSO disks and high-contrast imaging.

\subsection{Mid-IR laser combs}
Vasisht et al. and Michael et al. (these proceedings) present progress on developing key technologies to allow broadband heterodyne interferometry in the terahertz regime.  While heterodyne detection will not be competitive with direct detection for wavelengths shorter than 8$\mu$m, there are some possibilities for the mid-IR, both N and Q bands.  Note that there are some unexplored atmospheric windows, such as 40$\mu$m that could be accessible if PFI is built at a dry site, like near ALMA or the high Antarctic plateau.  Currently, it is crucial to develop the technology toolkit for mid-IR broadband heterodyne, this includes laser combs, high-speed detector arrays, and the high-speed digital infrastucture for data processing.

\acknowledgments 
Members of the PFI SWG/TWG would like to thank our colleagues for supporting PFI and for many important suggestions and ideas over the past two years.  Interested colleagues should see planetformationimager.org for more information on the project, including on how to get more involved.
JDM would like to further acknowledge Professor Charles Townes for his profound impact on the field of stellar interferometry -- Dr. Townes passed away January 27, 2015 and will be missed by many in our field.  SPIE meetings will never be the same.


\bibliography{main} 

\bibliographystyle{spiebib} 

\end{document}